%% file: 0-camera_ready_aies24.tex
\documentclass[letterpaper]{article} 
\usepackage{aaai24}  
\usepackage{times}  
\usepackage{helvet}  
\usepackage{courier}  
\usepackage[hyphens]{url}  
\usepackage{graphicx} 
\usepackage{natbib}  
\usepackage{caption} 
\usepackage{algorithm}
\usepackage{algorithmic}
\usepackage{verbatim}
\usepackage{xcolor}

\usepackage{newfloat}
\usepackage{listings}
\DeclareCaptionStyle{ruled}{labelfont=normalfont,labelsep=colon,strut=off} 
\floatstyle{ruled}
\newfloat{listing}{tb}{lst}{}
\floatname{listing}{Listing}
\title{Trusting Your AI Agent Emotionally and Cognitively: Development and Validation of a Semantic Differential Scale for AI Trust}
\author {
    Ruoxi Shang\textsuperscript{\rm 1},
    Gary Hsieh\textsuperscript{\rm 1},
    Chirag Shah\textsuperscript{\rm 1}
}
\affiliations {
    \textsuperscript{\rm 1}University of Washington\\
    rxshang@uw.edu, garyhs@uw.edu, chirags@uw.edu
}

\usepackage{bibentry}

\begin{document}

\maketitle

\input{1-abstract.tex}

\input{2-introduction.tex}
\input{3-related-work-v2.tex}
\input{4-development.tex}

\input{5-development-results.tex}
\input{6-validation-a}

\input{7-validation-a-results}
\input{8-study-3.tex}
\input{9-study-3-results}
\input{10-discussion.tex}

\input{11-limitation.tex}

\bibliography{bibliography}

\section{Appendix}
\input{Appendix}
\end{document}

%% file: 1-abstract.tex
Trust is not just a cognitive issue but also an emotional one, yet the research in human-AI interactions has primarily focused on the cognitive route of trust development. Recent work has highlighted the importance of studying affective trust towards AI, especially in the context of emerging human-like LLMs-powered conversational agents. However, there is a lack of validated and generalizable measures for the two-dimensional construct of trust in AI agents. To address this gap, we developed and validated a set of 27-item semantic differential scales for affective and cognitive trust through a scenario-based survey study. We then further validated and applied the scale through an experiment study. Our empirical findings showed how the emotional and cognitive aspects of trust interact with each other and collectively shape a person's overall trust in AI agents. Our study methodology and findings also provide insights into the capability of the state-of-art LLMs to foster trust through different routes.


%% file: 2-introduction.tex
\section{Introduction} \label{sec:introduction}



Trust plays a crucial role not only in fostering cooperation, efficiency, and productivity in human relationships \cite{brainov1999contracting} but also is essential for the effective use and acceptance of computing and automated systems, including computers \cite{madsen2000measuring}, automation \cite{lee2004trust}, robots \cite{hancock2011meta}, and AI technologies \cite{kumar2021intelligent}, with a deficit in trust potentially causing rejection of these technologies \cite{glikson2020human}. The two-dimensional model of trust, encompassing both cognitive and affective dimensions proposed and studied in interpersonal relationship studies \cite{mcallister1995affect, johnson2005cognitive, parayitam2009interplay, morrow2004cognitive}, have been adopted in studying trust in human-computer interactions, particularly with human-like technologies \cite{hu2021dual, glikson2020human}. Cognitive trust relates to the perception of the ability (e.g., skills, knowledge, and competencies), reliability, and integrity of the trustee, whereas the affective dimension involves the perceived benevolence and disposition to do good of the trustee \cite{johnson2005cognitive, mayer1995integrative}. In the context of human-computer trust, cognition-based trust is built on the user's intellectual perceptions of the system's characteristics, whereas affect-based components are those which are based on the user's emotional responses to the system \cite{madsen2000measuring}.

While AI trust research has largely centered on technical reliability and competency, there is a notable lack of work that explores the affective routes of trust development. The recent advancement of text-based Large Language Models (LLMs) have demonstrated a remarkable ability to take on diverse personas and skill-sets, recognizing and responding to people's emotional needs during conversation-based interactions. This capability is crucially aligned with the increasing focus on simulating Affective Empathy in human-AI interactions \cite{paiva2017empathy, welivita2021large}. In light of this, there is growing research interest in studying affective aspects of trust in AI \cite{glikson2020human, granatyr2017need, kyung2022rationally, zhang2021you, guerdan2021toward}. However, a critical gap exists in the lack of generalizable and accurate specialized measurement tools for assessing affective trust in the context of AI, especially with the enhanced and nuanced capabilities of LLMs. This highlights a need for a better measurement scale for affective trust to gain a deeper understanding of how trust dynamics function, particularly in the context of emotionally intelligent AI.

In this paper, 
we introduce a 27-item semantic differential scale (See Table \ref{tab:items}) for assessing cognitive and affective trust in AI, aiding researchers and designers in understanding and improving human-AI interactions. Our motivation and scale development process is based on a long strand of prior research on the cognitive-affective construct of trust that has been shown to be important in interpersonal trust in organizations, human trust in conventional technology and automation, and more recently in trust towards AI. Our use of OpenAI's ChatGPT to generate different levels of affective trust further demonstrates a scalable method for studying the emotional pathway to AI trust. Empirically, we contribute findings on 
the interplay and distinction between cognitive, affective, and moral trust.
The paper is structured to highlight these contributions: Section \ref{sec:study-dev} describes the development and validation of our trust scale through an experimental survey study and factor analysis. The validation and application of our scale is described in Section \ref{sec:study-2}. Section \ref{sec:study-2-a} begins with a preliminary study validating scales and testing LLMs as a tool to manipulate affective trust. This is followed by a refined study \ref{sec:study-3-design} to further establish discriminant validity and explore cognitive-affective trust dynamics. Section \ref{sec:discussion} then discusses the implications of these findings as well as potential usage of our trust scale.


%% file: 3-related-work-v2.tex
\section{Related Work} \label{sec:related-work}

\subsection{Shifting Paradigm of AI Trust Research} \label{sec:related-work-sub1}
Due to the opaque nature of most high-performing AI models, trust between the user and the AI system has always been a critical issue \cite{carvalho2019machine, thiebes2021trustworthy, jacovi2021formalizing}, as inappropriate trust can lead to over-reliance or under-utilization of AI systems \cite{buccinca2021trust, asan2020artificial}. Research in trust has predominantly adopted the cognitive evaluation of the system's performance \cite{granatyr2015trust}, such as its accuracy in making predictions \cite{ribeiro2016should}, its perceived consistency in completing tasks \cite{mcknight2011trust}, and its ethical considerations and transparency in decision-making \cite{duran2021afraid}. 

Studies in psychology have long been establishing the importance of psychological influence (e.g., emotions, personality, moods) on trust \cite{dunn2005feeling, lount2010impact}. Extending beyond the traditional numeric and cognitive paradigm, recent works have proposed the importance of exploring affective factors of trust in AI systems \cite{granatyr2017need, gillath2021attachment, jeon2023effects}. For technologies perceived as more human-like, affective trust factors such as benevolence and integrity play a more significant role \cite{lankton2015technology}. Moreover, recent advancements in AI, particularly in Large Language Models (LLMs) has demonstrated its capability beyond traditional task performance, as scholars find it challenging not to anthropomorphize them \cite{shanahan2022talking}. Notably, OpenAI's GPT-4, has shown excellent performance in Emotional Awareness (i.e. the ability to identify and describe emotions) \cite{elyoseph2023chatgpt}. There is also increasing interest in studying LLMs' empathetic responses \cite{ayers2023comparing, belkhir2023beyond}. Our work extends the current focus on the emotional aspects of AI interactions by highlighting the need to explore the emotional dimension of trust, a concept with deep roots in research studying interpersonal relationships.

\subsection{Affective and Cognitive Trust} \label{sec:related-work-sub2}
The interdisciplinary nature of AI trust research motivates the adoption of theoretical frameworks from interpersonal relationship literature \cite{bansal2023workshop, thiebes2021trustworthy}. Among the classic interpersonal trust theories and models (e.g., \cite{mayer1995integrative, rempel1985trust}), a two-dimensional model with cognitive and affective components has been extensively studied \cite{mcallister1995affect}. Similar to trust towards humans, trust towards technology has both cognitive and affective components \cite{komiak2004understanding}. In the AI context, cognitive trust relates to the user's intellectual perceptions of the AI's characteristics \cite{komiak2004understanding, madsen2000measuring}, focusing on aspects like reliability and transparency. Affective trust, on the other hand, involves emotional responses to the AI, including factors like tangibility and anthropomorphism \cite{ueno2022trust, glikson2020human}. This duality is essential due to the inherent complexity of AI, which often suggests a need for a "leap of faith" in its hidden processes, beyond what can be cognitively processed \cite{hoff2015trust, lee2004trust}. Prior works have found the limitation of cognition in decision-making, as demonstrated by studies showing limitations in users' abilities to discern AI inaccuracies, even with support through explanations \cite{jacobs2021machine, buccinca2021trust}. The cognitive-affective architecture has been established in research of computational agents \cite{perez2016cognitive, chumkamon2016intelligent}. The importance of this bi-dimensional model lies in its capacity to capture the full spectrum of trust dynamics that single-dimensional models, focusing solely on either aspects, fail to encompass. Trust has also been investigated through other bi-dimensional models in Human-Robot Interaction (HRI) (e.g. Law and Scheutz's Performance-based and Relation-based trust \cite{law2021trust}, and Malle and Ullman's Multi-Dimensional Measure of Trust (MDMT) \cite{malle2021multidimensional}). Our work makes a unique contribution by focusing on the Cognitive-Affective (C-A) trust model that fully encapsulates the emotional and psychological intricacies in the interactions with the state-of-the-art AI models that have advanced emotional intelligence. Although MDMT was derived from a different body of prior literature in social-moral constructs as mentioned in their work \cite{ullman2018does, ullman2019measuring}, we found it to be a suitable scale to compare it with due to its similar bi-dimentional construct and adjective item format with our scale. Therefore, in Section~\ref{sec:study-3-design}, we use the moral scale to establish discriminant validity with our cognitive-affective trust scale, demonstrating the distinctiveness of our cognitive-affective trust scale.



\subsection{Role and Effects of Affective Trust}\label{sec:related-work-sub3}

There is growing research interest in exploring the role of affective trust in the use of AI technologies. A few recent works have highlighted that affect-based trust plays a decisive role in people's acceptance of AI-based technology in preventative health interventions \cite{kyung2022rationally} and financial services robo-advising \cite{zhang2021you}. Research in explainable AI (XAI) has also shown that people's affective responses to explanations are crucial in improving personalization and increasing trust in AI systems \cite{guerdan2021toward}. However, given the interdisciplinary nature of AI trust research, the valuable insights to be borrowed from interpersonal trust are currently understudied in the AI context. Prior work has found that affective and cognitive trust have different impacts on relationships \cite{webber2008development, mcallister1995affect}. Cognitive trust tends to form rapidly \cite{mcknight2002impact, meyerson1996swift}, whereas affective trust is more persistent under challenges in teamwork \cite{mcallister1995affect} and interpersonal relationships \cite{williams2001whom}. Affective trust also shows greater resilience to short-term issues and errors \cite{jones1998experience, mcallister1995affect}. Researchers have also shown that affective and cognitive trust are not isolated constructs; rather, they complement each other \cite{granatyr2017need}, and affective trust needs to be developed on the basis of cognitive trust \cite{johnson2005cognitive}. Acknowledging these research opporunities, our work is a step towards a deeper and holistic examinination of the complex dynamics between cognitive and affective trust and their contribution to general trust in AI.

\subsection{Gaps in Empirical Research and Measurement of Affective Trust in AI} \label{sec:related-work-sub4}
Despite growing interest in this space, existing studies and measurement scales for affective trust in AI exhibit limitations, particularly in the adaptation and validation of measurement scales. Many existing scales, primarily developed for human trust contexts, have been applied to AI interactions with minimal modifications, raising questions about their generalizability. For instance, trust items intended for Human-Computer Trust were directly used for AI systems handling personal data, without substantial revision to reflect the unique aspects of AI interactions \cite{liao2021should}. Furthermore, there's a lack of consensus on defining affective trust in AI. While Kyung and Kwon \cite{kyung2022rationally} merged benevolence and integrity dimensions to measure affective trust in AI-based health interventions, Shi et al. \cite{shi2021antecedents} categorized these dimensions as cognitive trust, employing a different scale \cite{komiak2006effects} for affective trust. This inconsistency highlights the need for a unified, valid measure of trust for AI technologies \cite{ueno2022trust}. Given the intertwined nature of affective and cognitive trust, it is evident that a comprehensive evaluation of trust in AI systems requires a scale that measures both dimensions. In response, this work adopts Verhagen et al.'s \cite{verhagen2015toward} approach, developing semantic differential scales for both affective and cognitive trust in AI. Unlike Likert-type scales, semantic differentials use bipolar adjective pairs, offering advantages in reducing acquiescence bias and improving robustness \cite{hawkins1974stapel}, reliability \cite{wirtz2003examination}, and validity \cite{van1995assessment}.

%% file: 4-development.tex
\section{Scale Development} \label{sec:study-dev}

\subsection{Initial Item Generation}
In developing our trust item pool, we conducted a comprehensive literature review to identify prominent two-dimensional trust models that differentiate between cognitive and affective components. We pooled models and items from literature in interpersonal trust, intraorganizational trust, and trust in interaction with computers and traditional technologies. This approach is consistent with the broader trend of extending trust research from human-human contexts to human-AI interactions, as evidenced by the comprehensive review by Glikson and Woolley \cite{glikson2020human}. After the initial literature review, we applied a rigorous selection process to a larger body of trust literature, using the following criteria: 1) clear focus and delineation of affective and cognitive dimensions; 3) wide citation and application in various contexts; and 4) applicability to human-AI interactions. This ensures a thorough representation of the most relevant frameworks. The final selected models include Lewis and Weigert's sociological model \cite{lewis1985trust}, McAllister's interpersonal trust model \cite{mcallister1995affect}, Madsen and Gregor's Human-Computer Trust Components \cite{madsen2000measuring}, Johnson and Grayson's customer trust model \cite{johnson2005cognitive}, and Komiak and Benbasat's IT adoption trust model \cite{komiak2006effects}. From these, we extracted 56 unique key adjectives from their scales. Subsequent refinement involved removing synonyms and ensuring coverage of key dimensions: reliability, predictability, competence, understandability, integrity, benevolence, and amiability, which were adopted from the subscales from the above-mentioned models. The dimensions are kept flexible and serves mainly as a reference for coverage. We also developed antonym pairs for each adjective using resources like Merriam-Webster and Oxford English Dictionary, selecting the most appropriate antonym after several review rounds among the researchers. This resulted in 33 paired adjective items, divided into cognitive ($N=20$) and affective ($N=13$) trust categories, as detailed in Table~\ref{tab:items}. In the following step, we recruited participants to rate these items with respect to various scenarios through an online survey study.

\subsection{Survey design}
\label{sec:study-1-survey-design}

We used the hypothetical scenario method, where participants evaluated vignettes describing realistic situations to rate trust-related scales \cite{trevino1992experimental}. This method is frequently used in studying trust in emerging or future-oriented intelligent systems \cite{shi2021antecedents, juravle2020trust, gillath2021attachment, kim2021you}. 
Hypothetical scenarios enable exploration of long-term, nuanced, human-like interactions with AI assistants. This method also facilitates control over variables like agent type and interaction types, and risk levels, ensuring generalizability. In addition, this method ensures consistency in contextual details across respondents \cite{alexander1978use}. We crafted 32 scenario variations, manipulating the following five key dimensions: \textbf{Trust Level} (high vs. low),\textbf{ Trust Route} (affective vs. cognitive),\textbf{ Prior Interaction }(first-time vs. repeated),\textbf{ Application Domain Stakes }(high vs. low), and \textbf{Agent Type} (human vs. AI).

For validation purpose of the scales, we manipulated \textbf{Trust Level} and \textbf{Trust Route}. This involved depicting the agent's characteristics and behaviors in the scenarios, aligning them with varying levels of cognitive or affective trust. Additionally, to ensure the scales' generalizability, we manipulated \textbf{Prior Interaction Frequency} to be interacting with the agent for the \textit{first time} or \textit{multiple times}, and we set \textbf{Application Domain Stakes} to be either high-stake domains (\textit{Healthcare Diagnostics} and \textit{Self-Driving Taxi}) and low-stake domains (\textit{Personal Training} and \textit{Study Tutor}), inspired by real-world applications. These manipulations were implemented through texts presented to participants, as illustrated in Figure \ref{fig:scenario-example} (See Appendix). The tests for the manipulation's effectiveness will be demonstrated in Section \ref{sec:dev-results-t-test}.


Each participant were presented with two text-based scenarios for repeated measures. A mixed-model experiment design was deliberately chosen to incorporate both within-subject and between-subject variables. \textbf{Agent Type} and \textbf{Prior Interaction} are set to vary within-subjects to capture nuanced differences despite individual variability, and \textbf{Application Domain Stakes} is also designed to vary within-subjects to prevent boredom from the repetition of content. The order in which they see the variations are randomized to control for order effect. The rest of the dimensions are between-subjects and randomly assigned to participants. The two scenarios in Figure \ref{fig:scenario-example} (See Appendix) showcase one of the possible pairs of scenarios a participant may encounter.

During the survey study, after being presented with the first text-based scenarios, participants were asked to rate the semantic differential adjective pairs on a five-step scale, as well as a question assessing general trust in the AI agent. This process is repeated for the second scenario. After completing both scenarios, participants responded to questions used for our control variables including AI literacy and demographic information. The scenario structure comprised two parts: a prompt setting the interaction context, and three sentences detailing the agent's characteristics and behaviors.

\subsection{Measurement and Variables}

In our survey, we evaluated several key variables. For \textbf{Affective and Cognitive trust}, we used our semantic differential scale, where participants rated 33 adjective antonym pairs on a scale of -2 (most negative) to 2 (most positive). \textbf{General trust} was measured using a single-item questionnaire adapted from Yin \cite{yin2019understanding}, where participants responded to the question "how much do you trust this AI assistant to provide you with the guidance and service you needed" on a 5-point Likert scale, ranging from 1 ("I don't trust this agent at all") to 5 ("I fully trust this AI"). \textbf{AI literacy} was assessed using items adapted from Wang \cite{wang2022measuring}, all rated on a 5-point Likert scale from "Strongly disagree" to "Strongly agree", including items like "I can identify the AI technology in the applications I use" and "I can choose the most appropriate AI application for a task"

\subsection{Participants}
Amazon Mechanical Turk (MTurk) has been frequently used to recruit participants for online scenario-based studies related to AI technologies \cite{antes2021exploring, kim2021you, kaur2020using}. We recruited 200 participants from the United States through Amazon Mechanical Turk. The eligibility criteria included a minimum of 10,000 HITs Approved and an overall HIT Approval Rate of at least 98\%. Each participant received a compensation of \$2.20. The study involved repeated measures, collecting two sets of responses per participant for the two scenarios. Our quality control measures included a time delay for scenario reading, four attention checks, exclusions for uniform ratings or completion times more than two standard deviations from the mean, and a randomized sequence to control for order effects. After applying these criteria, we excluded 49 participants, resulting in 151 valid responses for the final analysis.

%% file: 5-development-results.tex
\subsection{Results} \label{sec:study-dev-results}
\subsubsection{Exploratory Factor Analysis} \label{sec:study-dev-EFA}
To uncover the factor structure underlying the 33 trust items, we first verified the suitability of our data for factor analysis. Bartlett's Test of Sphericity showed significant results ($\chi^2=12574, p<0.001$) \cite{bartlett1950tests}, and the Kaiser-Meyer-Olkin Measure of Sampling Adequacy was high at 0.98 \cite{kaiser1970second, dziuban1974correlation}, both indicating the appropriateness of factor analysis for our dataset. To determine the number of trust sub-components, we applied Kaiser's eigenvalue analysis \cite{kaiser1958varimax} and parallel analysis \cite{hayton2004factor}, which collectively suggested a two-factor structure
.

We initially used an oblique rotation as recommended by Tabachnick and Fiddell for instances where factor correlations exceed 0.32 \cite{tabachnick2013using}. Given the high correlation among our factors ($r = 0.78$) \cite{gorsuch1988exploratory}, we retained this rotation method. We then refined our item pool based on specific criteria: items were kept only if they had a factor loading above 0.4 \cite{howard2016review}, ensuring significant association with the underlying factor. Items with a cross-loading of 0.3 or more were removed to align item responses with changes in the associated factor \cite{howard2016review}. Additionally, we applied Saucier's criterion, eliminating items unless their factor loading was at least twice as high as on any other factor \cite{saucier1994mini}. This led to the removal of six items: Harmful - Well-intentioned, Unpromising - Promising, Malicious - Benevolent, Discouraging - Supportive, Insincere - Sincere, and Unpleasant - Likable.


A second round of exploratory factor analysis with the remaining 27 items preserved all items, as they met the above-mentioned criteria. The final item loadings are presented in Table \ref{tab:items} under the ``All" column, with empty rows indicating the eliminated items. All remaining items demonstrated primary loadings above $0.55$. Upon examining the keywords of items in each factor, two distinct themes emerged: cognitive trust and affective trust. This alignment was consistent with the dimensions identified in the initial literature review. Factor 1, representing cognitive trust, accounted for 43\% of the total variance with 18 items, while Factor 2, corresponding to affective trust, explained 23\% with 9 items.


\subsubsection{Reliability}
To test the internal reliability of the resulting items, we computed Cronbach's $\alpha$ for each scale. The cognitive trust scale ($\alpha = .98)$ and the affective trust scale ($\alpha = .96$) both showed high internal consistency. We also tested the item-total correlation between each item and the average of all other items in the same sub-scale. All items' correlations exceed 0.6. In this development study, 18 items measuring cognitive trust and 9 items measuring affective trust were identified with high reliability.

\subsubsection{Construct Validity} \label{sec:dev-results-t-test}
In addition to high reliability, we conducted analyses to show the validity of our scale. We first examined the construct validity, which refers to the degree to which the scale reflects the underlying construct of interest. Recall that we manipulated affective trust and cognitive trust through the level of trustworthiness and the trust development routes and controlled for factors like agent type, interaction stage, and risk level. T-test results revealed significant distinctions in both affective and cognitive trust scales under the experiment manipulation. Cognitive trust scale demonstrated a pronounced difference in high versus low cognitive trust conditions ($t=45.74, p<0.001$), and affective trust scale also showed a pronounced disparity in high versus low affective trust conditions ($t=43.00, p<0.001$). This also demonstrates the efficacy of our manipulation with the scenarios, as we observed significant differences in both the cognitive and affective dimensions.


We then fitted two separate linear random effect models \cite{singmann2019introduction} on the two scales over the two manipulations due to our experiment design. Model 1 and Model 2 in Table \ref{tab:dev-regression-aff-cog} (See Appendix) tests the effects of our manipulations on the resulting trust scales, while Model 3 tests the effects of both scales on general trust. As shown in Table \ref{tab:dev-regression-aff-cog} (See Appendix), we observed significant main effects of manipulation \textbf{Trust Level} ($r = 2.059, p<0.001$) and manipulation \textbf{Trust Route} ($r = -0.497, p<0.01$) of these two manipulations on the cognitive trust scale, and the same is observed for affective trust scale. More importantly, the interaction effect shows that the affective trust scale is higher when higher trust is developed via the affective route ($r=0.921, p<0.001$), while the cognitive trust scale is higher when higher trust is developed via the cognitive route ($r=-0.538, p<0.05$). The above analyses demonstrated the construct validity of our scale.

\subsubsection{Concurrent Validity}

We then examined concurrent validity that assesses the degree to which a measure correlates with a establish criterion, which is a single-item measuring general trust towards the agent. After confirming that general trust for the agent was significantly higher in the higher trustworthiness condition ($t = 10.47, p < 0.001$), we found that overall trust is significantly and positively predicted by both the cognitive trust scale ($r=0.881, p<0.001$) and the affective trust scale ($r=0.253, p<0.001$). The effect size of the cognitive trust scale on general trust is greater than that of the affective trust scale. This is also consistent with the previous factor analysis result that the cognitive trust scale explains more variance than the affective trust scale. These convergent tests provided sufficient support for the validity of our scales. Hence, in the next step, we applied them to measuring cognitive and affective trust in conversational AI agents. 

\input{tables/table-1-items-all}

%% file: tables/table-1-items-all.tex
\begin{table*}[]
\caption{This table presents 33 initial cognitive and affective trust items as antonym pairs. \textbf{Items in black represent the final scale with 27 items retained after exploratory factor analysis (EFA), while grey items were eliminated (shown with empty factor loadings).} Factor loadings are shown in columns F1 and F2 for both the complete dataset ('All') and AI agent condition subset, demonstrating consistent two-factor structure across both analyses. Sources are indicated by letters: a) \cite{mcallister1995affect}, b) \cite{johnson2005cognitive}, c) \cite{erdem2003cognitive}, d) \cite{madsen2000measuring}, e) \cite{gefen2002reflections}, f) \cite{komiak2006effects}.}
\label{tab:items}
\begin{tabular}{llllllll}
 &
   &
   &
  All &
   &
  AI Condition &
   &
   \\ \hline
\# &
  Item &
  Sub-dimension &
  F1 &
  F2 &
  F1 &
  F2 &
  Source \\ \hline
C1 &
  Unreliable - Reliable &
  Reliability &
  0.905 &
  0 &
  0.922 &
  -0.058 &
  a,b,c,d,e \\
C2 &
  Inconsistent - Consistent &
   &
  0.928 &
  -0.12 &
  0.924 &
  -0.183 &
  a,b,c,d \\
C3 &
  Unpredictable - Predictable &
   &
  0.894 &
  -0.171 &
  0.898 &
  -0.204 &
  d \\
C4 &
  Undependable - Dependable &
   &
  0.851 &
  0.016 &
  0.911 &
  -0.111 &
  a,b,c,d \\
C5 &
  Fickle - Dedicated &
   &
  0.759 &
  0.139 &
  0.744 &
  0.116 &
  a,b,d \\
C6 &
  Careless - Careful &
   &
  0.721 &
  0.213 &
  0.716 &
  0.206 &
  a,b \\
C7 &
  Unbelievable - Believable &
   &
  0.69 &
  0.082 &
  0.658 &
  0.034 &
  c,d,e \\
C8 &
  \textcolor{lightgray}{Unpromising - Promising} &
   &
   &
   &
   &
   &
  c,d,e \\
C9 &
  Clueless - Knowledgable &
  Competence &
  0.907 &
  -0.018 &
  0.898 &
  0.026 &
  f,d,e \\
C10 &
  Incompetent - Competent &
   &
  0.9 &
  0.023 &
  0.921 &
  -0.021 &
  a,b,f,d,e \\
C11 &
  Ineffective - Effective &
   &
  0.861 &
  0.075 &
  0.863 &
  0.071 &
  d,e \\
C12 &
  Inexperienced - Experienced &
   &
  0.751 &
  0.089 &
  0.611 &
  0.155 &
  a,b,f,e \\
C13 &
  Amateur - Proficient &
  Understandability &
  0.895 &
  0.009 &
  0.864 &
  0.039 &
  a,b,c,f,d,e \\
C14 &
  Irrational - Rational &
   &
  0.827 &
  0.02 &
  0.792 &
  0.05 &
  d,e \\
C15 &
  Unreasonable - Reasonable &
   &
  0.71 &
  0.224 &
  0.714 &
  0.202 &
  d,e \\
C16 &
  Incomprehensible - Understandable &
   &
  0.706 &
  0.175 &
  0.783 &
  0.079 &
  f,d \\
C17 &
  Opaque - Transparent &
   &
  0.6 &
  0.261 &
  0.656 &
  0.18 &
  c,f,d \\
C18 &
  Dishonest - Honest &
  Integrity &
  0.693 &
  0.178 &
  0.743 &
  0.097 &
  c,f,e \\
C19 &
  Unfair - Fair &
   &
  0.663 &
  0.274 &
  0.66 &
  0.268 &
  c,f \\
C20 &
  \textcolor{lightgray}{Insincere - Sincere} &
   &
   &
   &
   &
   &
  f,e \\ \hline
A1 &
  Apathetic - Empathetic &
  Benevolence &
  -0.11 &
  0.989 &
  -0.162 &
  0.967 &
  a,b \\
A2 &
  Insensitive - Sensitive &
   &
  -0.08 &
  0.959 &
  -0.109 &
  0.955 &
  a,b,c \\
A3 &
  Impersonal - Personal &
   &
  -0.024 &
  0.902 &
  -0.025 &
  0.847 &
  b,c,d \\
A4 &
  Ignoring - Caring &
   &
  0.048 &
  0.881 &
  -0.055 &
  0.941 &
  a,b,c \\
A5 &
  Self-serving - Altruistic &
   &
  0.215 &
  0.627 &
  0.207 &
  0.622 &
  b,c,e \\
A6 &
  \textcolor{lightgray}{Malicious - Benevolent} &
   &
   &
   &
   &
   &
  f,e \\
A7 &
  \textcolor{lightgray}{Harmful - Well-intentioned} &
   &
   &
   &
   &
   &
  f,e \\
A8 &
  \textcolor{lightgray}{Discouraging - Supportive} &
   &
   &
   &
   &
   &
  c,f,e \\
A9 &
  Rude - Cordial &
  Amiability &
  -0.11 &
  0.989 &
  0.112 &
  0.76 &
  a,b \\
A10 &
  Indifferent - Responsive &
   &
  0.221 &
  0.711 &
  0.232 &
  0.667 &
  a,b,c,f,e \\
A11 &
  Judgemental - Open-minded &
   &
  0.142 &
  0.688 &
  0.078 &
  0.697 &
  c \\
A12 &
  Impatient - Patient &
   &
  0.291 &
  0.577 &
  0.218 &
  0.6 &
  a,b \\
A13 &
  \textcolor{lightgray}{Unpleasant - Likable} &
   &
   &
   &
   &
   &
  f,d
\end{tabular}
\end{table*}

%% file: 6-validation-a.tex
\section{Scale Validation} \label{sec:study-2}
After developing a reliable two-factor scale for measuring cognitive and affective trust in AI, we conducted two validation studies. Study A (Section \ref{sec:study-2-a}) validated the scale using Confirmatory Factor Analysis and tested the efficacy using LLM-generated conversations to elicit different levels of trust. Building on Study A's findings, Study B (Section \ref{sec:study-3-design}) refined the study design to establish discriminant validity of the scales and provide empirical insights into the interaction between the two trust dimensions.

\subsection{Validation Study A - Preliminary Study} \label{sec:study-2-a}
In this study, in addition to establish scale validity, we test the efficacy of our affective trust scale in distinguishing between two conversational AI assistants with mock dialogues generated by OpenAI's ChatGPT \cite{chatgpt}, a leading example of state-of-the-art LLM-based conversational agents. We used pre-generated mock-up conversations to reduce variations and errors induced in the interaction with LLMs, controlling for the effect of our manipulation. This survey study was initiated with uncertainties regarding GPT models' ability to evoke varying degrees of affective trust. Hence, we conducted a preliminary study to assess the effectiveness of ChatGPT and the sensitivity of our scale to the applied manipulations.

\subsubsection{Study Design and Participants}
We designed a within-subjects online experiment, in which participants evaluated screenshots of dialogues with two AI assistants, \textit{Echo} and \textit{Nova} (See Appendix for examples). \textit{Echo} was designed to elicit high affective trust, while \textit{Nova} demonstrated a lack of it. Our hypotheses were: affective trust would be higher for \textit{Nova} than \textit{Echo} (H1), and based on previously observed correlation between affective and cognitive trust, cognitive trust would also be higher for \textit{Nova} (H2).


To explore the feasibility and efficacy of Large Language Models (LLMs) in manipulating affective trust, we used ChatGPT to generate AI responses, leveraging its capability for human-like interactions to manipulate affective trust levels and at the same time controlling for the speech style and length. After validating the definitions of affective and cognitive trust generated by ChatGPT against literature, we crafted prompts to vary affective trust levels. After experimenting with different prompts and scenarios, we chosen the scenario of user asking the AI agent for emotional support, in which the user starts with the question ``Lately, I've been feeling lonely. What should I do?" The responses were generated by ChatGPT and lightly edited for conciseness.

In addition to measuring affective and cognitive trust with our 27-item developed scale, we also included disposition to trust, AI literacy, age, and gender were included as control variables because previous studies have demonstrated their impacts on trust \cite{shi2021antecedents}. AI familiarity was measured by 3 survey questions including "I am familiar with using an AI-powered chatbot to help me with specific tasks" on a 7-point Likert scale. AI literacy is measured by the same items as in the previous survey. Trust Disposition was measured by items adopted from prior work \cite{gefen2000commerce}. General trust in each chatbot was measured using a one-item scale adapted from prior research \cite{ueno2022trust}.


We conducted our experiment via Amazon MTurk, where participants viewed two screenshots, each depicting a three-question conversation with either the AI chatbot Echo or Nova. After viewing each conversation, they rated them using the semantic differential scales developed in our previous study. To avoid order effects, the sequence of viewing Echo and Nova was randomized. Post-assessment, they completed additional questions on trust disposition, AI literacy, and demographics. Following the same protocol of our development study, we recruited and filtered the data, ultimately analyzing 44 out of 50 participants' responses. A total of 88 responses were included in the final analysis due to repeated measures.



%% file: 7-validation-a-results.tex
\subsection{Results}

\subsubsection{t-tests}
\label{sec:study-2-results}
Welch's t-tests showed that general trust ($t = 2.37, p < 0.05$), affective trust scale ($t=3.78, p<0.001$), and cognitive trust scale ($t=2.84, p<0.01$) all yielded significant differences between high and low affective trust conditions. This shows that the manipulation using ChatGPT is successful. ChatGPT has the capability of eliciting different levels of affective trust based on the model’s learned representation of affective trust.

\subsubsection{Confirmatory Factor Analysis}
To confirm the factor structure determined by the EFA and assess its goodness-of-fit compared to alternative models, we performed a Confirmatory Factor Analysis (CFA) \cite{hinkin1997scale, long1983confirmatory}, which is a structural-equations analysis that compares the fit of rival models. We conducted Confirmatory Factor Analysis (CFA) using both Maximum Likelihood (ML) and Diagonally Weighted Least Squares (DWLS) estimators to assess the fit of our model against a one-factor baseline model.  We calculated several goodness-of-fit metrics, including the Comparative Fit Index (CFI), which measures the model's fit relative to a more restrictive baseline model \cite{bentler1990comparative}; the Tucker-Lewis Index (TLI), a more conservative version of CFI, penalizing overly complex models \cite{bentler1980significance}; the Standardized Root Mean Square Residual (SRMR), an absolute measure of fit calculating the difference between observed and predicted correlation \cite{hu1999cutoff}; and the Root Mean Square Error of Approximation (RMSEA), which measures how well the model reproduces item covariances, instead of a baseline model comparison \cite{maccallum1996power}. The ML estimator yielded mixed results, with some fit indices suggesting adequate fit (CFI = 0.920, TLI = 0.914, SRMR = 0.046) while others indicated suboptimal fit (RMSEA = 0.082, $\chi^2(494) = 1506.171$, $p < 0.001$). However, when using the DWLS estimator, which is more appropriate for our ordinal data \cite{li2016confirmatory, mindrila2010maximum}, the model demonstrated excellent fit across all indices (CFI = 1.000, TLI = 1.003, RMSEA = 0.000, SRMR = 0.038, $\chi^2(494) = 250.936$, $p = 1.000$). Robust fit indices, which account for non-normality in the data, also supported the model's fit using both estimators (ML: Robust CFI = 0.941, Robust TLI = 0.937, Robust RMSEA = 0.071; DWLS: Robust CFI = 0.998, Robust TLI = 0.997, Robust RMSEA = 0.035). The model fit significantly better than the baseline model using both ML ($\chi^2(528) = 13132.525$, $p < 0.001$) and DWLS ($\chi^2(528) = 78914.753$, $p < 0.001$) estimators. Overall, these results provide strong evidence for the adequacy of our proposed factor structure, particularly when using estimators well-suited for ordinal data.

\subsubsection{Validity tests}
We examined construct validity followed by concurrent validity of our scale following the same procedure as in the previous study. We first tested construct validity by checking the two scales are sensitive to the manipulation of affective trust through three regression models as shown in Table \ref{tab:val-regression-aff-cog} (See Appendix). 
Model 1 and Model 2 test the effects of our manipulations on the affective and cognitive trust scales respectively. Model 3 tests the effects of both scales on general trust. We observed the main effects of the condition on both affective and cognitive trust scales. Interacting with an AI chatbot with higher affective trustworthiness led to 0.95 points higher on the 7-point affective scale and the cognitive trust scale was increased by 0.80 points. This differential impact highlights the scale's nuanced sensitivity: while both affective and cognitive trusts are influenced by affective trust manipulation, the affective trust scale responded more robustly. Concurrent validity was then affirmed through significant positive predictions of general trust by both the affective trust scale ($r = 0.486, p < 0.001$) and the cognitive trust scale ($r = 0.546, p < 0.001$).



%% file: 8-study-3.tex
\subsection{Validation Study B - Refined Study}\label{sec:study-3-design} 

The preliminary study established the practical validity of our AI trust scale and demonstrating the effectiveness of using ChatGPT to manipulate affective trust.  It also provides empirical support for the scale's sensitivity to variations in trust levels induced by different attributes of an AI agent's communication style. Building on this foundation, this study aimed to delve deeper into the interplay between affective and cognitive trust, while also comparing our scale with the Multi-Dimensional Measure of Trust (MDMT) to establish discriminant validity. This comparative analysis sought to highlight the distinctiveness of our affective trust scale and the importance of establishing it as a separate scale.

We chose the Moral Trust Scale from Multi-Dimensional Measure of Trust (MDMT) model for a comparative analysis with our developed affective trust scale for AI, primarily due to its established reputation in HRI research \cite{malle2021multidimensional, ullman2019measuring}, as mentioned previously in Section \ref{sec:related-work-sub2}. Aside from both ours and MDMT being a two-dimensional trust models, our cognitive trust scale aligns closely with MDMT's capability trust scale, with overlapping scale items. This raises the question of whether our affective trust scale is measuring the same underlying construct as MDMT's moral trust scale. This comparison is crucial in highlighting the distinctiveness and specificity of our scale, particularly in capturing affective nuances in AI interactions that the moral trust might not cover. 

The findings from the preliminary laid the groundwork for the more complex experimental designs in this study. This study refined the previous design into a 2x2 fully-crossed factorial model with between-subject design, contrasting high and low levels of affective and cognitive trust. Multi-turn Q\&A conversations in each scenario were used to more effectively shape trust perceptions. We introduced two distinct scenarios: one involving Wi-Fi connectivity (primarily invoke cognitive trust) and another on handling interpersonal conflicts (primarily invoke affective trust). The two scenarios, each leaning more towards one aspect of trust, ensure that participants were not overly exposed to one type of trust over the other. This scenarios chosen represent everyday situations that are relatable for participants to ensure generalizability of our findings.

Similar to the previous study, we prompted ChatGPT to generate responses that are aim to elicit different levels of cognitive and affective trust by including or excluding elements related to these two different trust routes. Participants were randomly assigned to one of four conditions: high in both affective and cognitive trust (HH), low affective/high cognitive (LH), high affective/low cognitive (HL), or low in both (LL). Each condition included the two scenarios, with the order of presentation and item responses counterbalanced to control for order effects. The rest of the survey design mirrored Study A. After reading the scenarios, participants rated items from the affective, cognitive, and MDMT moral trust scales on a semantic differential scale from $-3$ to $+3$. They then assessed their general trust level towards the AI on a scale of $1$ to $7$. Following these ratings, we also collected additional data including trust disposition, AI literacy, AI familiarity, age, education level.

We recruited 180 participants on Prolific, presenting them with two ChatGPT conversations and the questions hosted on a Qualtrics survey form. Following the same quality control protocols as the previous studies, 168 responses were used int the final analysis.


%% file: 9-study-3-results.tex
\subsection{Results}
\label{sec:study-3-results} 

\subsubsection{t-tests for Manipulation Check}
We first conducted Welch's t-tests to check the effects of our experimental manipulations on the scale ratings. The conditions, categorized as High and Low, were designed to elicit the levels of cognitive and affective trust. Significant variations were noted in the affective trust scale between high and low affective trust conditions ($t=7.999, p<0.001$), and similarly in the cognitive trust scale between high and low cognitive trust conditions ($t=9.823, p<0.001$). These findings confirm the effectiveness of the manipulation.

\subsubsection{Factor Analysis}
We conducted exploratory factor analysis (EFA) to confirm the distinctiveness of scales, not for refactoring previously developed scales. The high Kaiser-Meyer-Olkin (KMO) value of $0.9597$ and a significant Bartlett's Test of Sphericity ($Chi-square=7146.38, p<0.001$) established the dataset's suitability for factor analysis. Three factors were retained, accounting for $70\%$ of the cumulative variance, a threshold indicating an adequate number of factors. This was also substantiated by a noticeable variance drop after the second or third factor in the scree plot and parallel analysis, where the first three actual eigenvalues surpassed those from random data. The results showed that the first three eigenvalues from our dataset were larger than the corresponding eigenvalues from the random data, indicating that these factors are more meaningful than what would be expected by chance alone. These results affirm that the items meaningfully load onto three distinct factors.

Our analysis used a factor loading threshold of 0.5 for clear factor distinctiveness. As shown in Table \ref{tab:factor_loadings} (See Appendix), EFA resulted in two main factors aligned with cognitive and affective trust scales, and a third factor predominantly linked to the Moral Decision-Making Trust (MDMT) scale, particularly its Ethical (Ethical, Principled, Has Integrity) and Sincere (Authentic, Candid) subscales. Items on MDMT's scale showed lower factor loadings in the same analysis, particularly in the emotional dimension, suggesting a weaker representation of affective elements. These outcomes underscore the distinct nature of the MDMT scale from the affective trust scale. Despite the overall clear conceptual distinction, we noted that the MDMT's ``Sincere" item and several cognitive trust items (Rational, Consistent, Predictable, Understandable, Careful, Believable) showed overlap across factors. This could be attributed to our study's design, which exclusively incorporates scenarios tailored to elicit affective and cognitive trust. This design choice was made to specifically examine these two types of trust, and also served as a way to determine if the moral trust scale reflects similar elements or different trust aspects not pertinent to our scenarios.

\subsubsection{Regression Analysis}
\label{sec:study-3-regression} 
We conducted regression analysis to compare the predictive power of the scales on general trust. Table \ref{tab:study-3-reg-a} (See Appendix) details this: Model 1 examines the effects of all three scales on general trust; Model 2 considers only cognitive and affective trust scales; and Model 3 includes the moral trust scale, excluding affective trust. This approach allows for comparison of the two related scales' contributions to general trust, while controlling for manipulation and other variables to observe in-group effects.


The results showed distinct contributions of each scale to general trust. Affective trust was a significant predictor in Model 1 ($r = 0.364, p < 0.01$) and Model 2 ($r = 0.376, p < 0.01$), whereas the moral trust scale showed non-significant correlations in all models. This suggests its limited relevance in scenarios dominated by emotional and cognitive cues. In contrast, the affective trust scale's significant impact highlights its ability to capture trust dimensions not addressed by the moral trust scale, demonstrating their distinctiveness. Additionally, among all the control variables that demonstrated significant impacts, AI familiarity positively influenced general trust in all models (Model 1: $r = 0.208, p<0.01$; Model 2: $r=0.209, p<0.01$; Model 3: $r=0.180, p<0.05$), whereas AI literacy negatively impacted trust in Model 1 ($r = -0.133, p<0.05$) and Model 2 ($r = -0.134, p<0.05$).

While affective and cognitive trust individually contribute to general trust, their interplay, particularly in conditions of imbalance, might reveal another layer of trust dynamics. We further explored the interaction between affective and cognitive trust in influencing general trust. As shown in Table \ref{tab:study-3-reg-b} (See Appendix), Models 1 and 2 showed no significant interaction effects with only cognitive trust scale showing strong, significant correlations (Model 1: $r = 0.799, p<0.001$; Model 2: $r = 0.849, p<0.001$). Model 3, however, revealed a significant negative interaction effect between high affective ($r = 1.677, p<0.001$) and cognitive trust ($r = 2.729, p<0.001$) conditions, despite their individual positive impacts. Figure \ref{fig:interaction-effect} (See Appendix) visually illustrates that when cognitive trust is high, changing affective trust has little effect on general trust. In contrast, under conditions of low cognitive trust, manipulating affective trust significantly impacts general trust. This means high cognitive trust overshadows the impact of the affective route on general trust, whereas low cognitive amplifies it.

%% file: 10-discussion.tex
\section{Discussion}
\label{sec:discussion}

\subsection{Scale Development and Validation}
Our work is grounded in the recognition that developing alternative instruments of established theoretical constructs holds significant value \cite{straub2004validation}. In this paper, we develop a validated affective trust scale for human-AI interaction and demonstrate its effectiveness at measuring trust development through the emotional route. While prior studies in AI trust have largely focused on cognitive trust, recent research emphasizes the need to consider affective trust in AI \cite{glikson2020human, granatyr2017need}. Existing affective trust scales, borrowed from models in non-AI contexts like interpersonal relationships and traditional computing \cite{mcallister1995affect, komiak2006effects}, lack rigorous validation for AI systems. Thus, our study develops and validates a scale for measuring both affective and cognitive trust in AI. Through a comprehensive survey study design and rigorous EFA process, we landed at a 18-item scale measuring cognitive trust and a 9-item scale measuring affective trust. The process resulted in the removal of six antonym pairs due to cross-loading, indicating their relevance to both trust dimensions. Through rigorous validation processes, we affirmed its reliability, internal consistency, construct validity, and concurrent validity.


The validation of our scales were carried out through two studies. In Study 2A, our analysis further demonstrates validity of the scale through CFA and a few follow-up validity tests. In Study 2B, the three-factor structure that emerged from the factor analysis, coupled with the insignificant coefficient from the regression analysis, provides clear evidence of discriminant validity. This indicates that our affective trust scale captures a distinct construct, separate from related scales measuring trust, such as the Multi-Dimensional Measure of Trust (MDMT) \cite{malle2021multidimensional}. The construction of our affective trust scale is key to this distinction; it includes a broader range of items that capture emotional nuances more effectively, thereby more accurately reflecting the affective pathway's impact on general trust. In contrast, MDMT's moral trust scale focuses on ethical (n=4) and sincerity (n=4) aspects. Some items in the sincerity subscale (e.g., sincerity, genuineness, candidness, authenticity) overlap with benevolence elements in our affective trust scale. However, our scale incorporates unique items like `Empathetic' and '`Caring,' absent in MDMT's scale, as well as likability aspects through items such as `Patient' and `Cordial.' These likability items are derived from established affective trust measures in human interactions, with previous studies confirming likability's role in fostering trust in various contexts including interpersonal relationships \cite{fiske2007universal}, digital platforms \cite{tran2023exploring}, and robot interactions \cite{cameron2021effect}. While they're distinct, we have also observed the relatively high correlation between these constructs. This empirical finding provide a valuable contribution to future work trying to reconcile these two different strands of research on trust.

Our final 27 item scale offers an adaptable tool for diverse research contexts and languages. Its simplicity, featuring just two adjectives per item, contrasts with the often context-specific declarative statements in Likert scales \cite{bruhlmann2020trustdiff}. This semantic differential format not only maintains reliability and validity during adaptation, but also usually leads to quicker survey completion compared to Likert scales \cite{chin2008fast}, facilitating widespread application to understand trust in AI technology. Developed through 32 scenarios across five dimensions and tested in two separate studies using everyday scenarios, the scale's generalizability extends to various domains and interaction durations with both human and AI assistants, making it versatile for future research comparing human and AI trust.

\subsection{Empirical Findings}
\subsubsection{LLMs-powered Tools as a Testing Bed}
With its proficiency in generating human-like responses, tools powered by LLMs such as ChatGPT stand out as a novel approach for examining trust in AI. This method significantly lowers the barriers to studying AI systems with emotional capabilities, particularly in manipulating trust via emotional routes. In our study, we found that GPT models' advanced conceptual understanding of affective and cognitive trust allows it to generate responses tailored to specific trust levels. This was demonstrated in our study (Section \ref{sec:study-3-results}). Our studies showed that LLMs effectively manipulate trust via cognitive and affective routes in diverse contexts like emotional support, technical aid, and social planning. This shows LLMs' versatility and utility in expediting trust formations in experimental studies. Our studies utilized pre-generated conversations to ensure control and consistency. Future research could explore the development of trust through LLMs in a different study setting, such as an interactive study setting or a longitudinal study setting with deeper relationship building.

\subsubsection{Interplay between Affective and Cognitive Trust}
Although previous research has established that affective and cognitive trust are distinct both conceptually and functionally \cite{mcallister1995affect, yang2009supervisory, johnson2005cognitive, zhu2014transformational}, our studies revealed a significant correlation between these two trust scales, echoing findings in prior work (e.g., \cite{de2016trust}). This indicates that while affective and cognitive trust are individual pathways to fostering trust, they are not isolated mechanisms and indeed influence each other. In addition, we identified a notable interaction effect between these two dimensions in shaping general trust in AI, as detailed in Section \ref{sec:study-3-regression} . When cognitive trust in the AI is already high, further manipulating affective trust does not significantly change overall trust. In contrast, when cognitive trust in a system is not high, influencing trust through emotional routes can be particularly helpful. This result aligns with prior work's finding in interpersonal relationship that affective trust often builds upon a foundation of cognitive trust \cite{johnson2005cognitive}. 

This finding of interaction effect highlight the potential for trust calibration \cite{zhang2020effect} in AI systems, particularly in contexts where cognitive trust is limited. This might arise during interactions with users having low literacy in AI \cite{long2020ai} and difficulty in achieving transparency, as with made even more challenging with LLMs \cite{liao2023ai}. Moreover, amidst the stochastic and occasionally unpredictable behavior of many AI systems, prior work has highlighted the affective route as trust repair strategies in designing trust resilient systems that despite occasional errors, remain fundamentally reliable and effective \cite{fahim2021integral}. However, it is crucial to note the risks of overtrusting AI through affective routes such as their social capabilities \cite{ullrich2021development}, and the potential for deceptive practices through the improper use of emotional communication \cite{coeckelbergh2011emotional}. Leveraging affective means to build trust is advocated only for AI systems that inherently possess cognitively trustworthy qualities, such as reliability and accuracy. For these AI systems, the emotional route can serve as a complementary approach to calibrate trust, especially when cognitive routes are less feasible.

\subsection{Potential Usage}
Our affective and cognitive trust scales present a valuable measurement tool for future research in designing trustworthy AI systems. Here, we outline a few possible usages. 

\subsubsection{Measure trust in human-AI interactions} 
The construct of trust with affective and cognitive dimensions is well-established in interpersonal trust literature. Our scale bridges the gap between human-human and human-AI trust, enabling future work to study trust in human-AI teaming to improve collaboration experiences and outcomes. For instance, our scale can be employed to investigate how these trust dimensions impact creative work with generative AI tools, as they have been found to influence team contributions differently \cite{ng2006contribute}. Furthermore, researchers have discovered that affective trust becomes more important later in the human teaming experience, while cognitive trust is crucial initially \cite{webber2008development}. Our scale offers the opportunity to examine the dynamics of these trust dimensions in human-AI collaboration.

\subsubsection{Support designing emotionally trustworthy systems}
Our research supports the growing understanding that emotional factors like empathy, tone, and personalization are crucial in establishing trust, especially in contexts where it's challenging to convey a system's performance and decision-making processes \cite{gillath2021attachment, kyung2022rationally}. Our scale can be used to distinctively measure trust developed through the affective route. This is particularly relevant in mental health interventions involving AI assistants, where patients may struggle to assess the AI's capabilities rationally \cite{hall2001trust, gillath2021attachment}. Affective trust becomes vital here, as patients, especially those with low AI literacy or experiencing anxiety, depression, or trauma, may respond more to emotional cues from AI, which typically lacks the emotional intelligence of human therapists. Our validated affective trust scale can guide the design of AI systems to calibrate for appropriate trust in this context, such as through empathetic responses or affect-driven explanations, and help explore its impact on long-term engagement and treatment adherence.

%% file: 11-limitation.tex
\section{Limitations and Future Work}

In our scale development phase, we designed scenario featuring AI agents as service providers. This role is chosen intentionally to align with prior affective trust research for interpersonal relationships \cite{johnson2005cognitive, komiak2004understanding}. Also, the prevalence of service-providing scenarios make it easier for general public participants to draw parallels between these AI agents with their human counterparts. 
Future work can explore other roles of AI, such as teammates \cite{zhang2023trust, zhang2021ideal} and friends \cite{brandtzaeg2022my}.

While our approach to categorizing trust dimensions into cognitive and affective aspects was informed by established trust frameworks (refer to Table \ref{tab:items}), the anticipated distinct subdimensions (e.g. reliability, understandability, etc.) were not as clear-cut after conducting exploratory factor analysis. This was possibly due to the subdimensions lacking sufficient unique variance or being highly correlated. Our scenario was deliberately designed to focused on differentiating cognitive and affective trust, while they might not have enough detailed information capture the nuances across the six dimensions. Future research to refine these subdimensions under cognitive and affective trust and examine their unique contributions to trust.



%% file: Appendix.tex
\input{figures/figure-1-scenarios}
\input{tables/table-study-1-results}

\input{figures/figure-study-2-mockup}
\input{tables/table-study-2a-results}

\input{tables/table-study-2b-main}
\input{tables/table-study-2b-int}
\input{figures/figure-2-int-and-factor-loadings}

%% file: figures/figure-1-scenarios.tex
\begin{figure*}
    \centering
    \includegraphics[width=\linewidth]{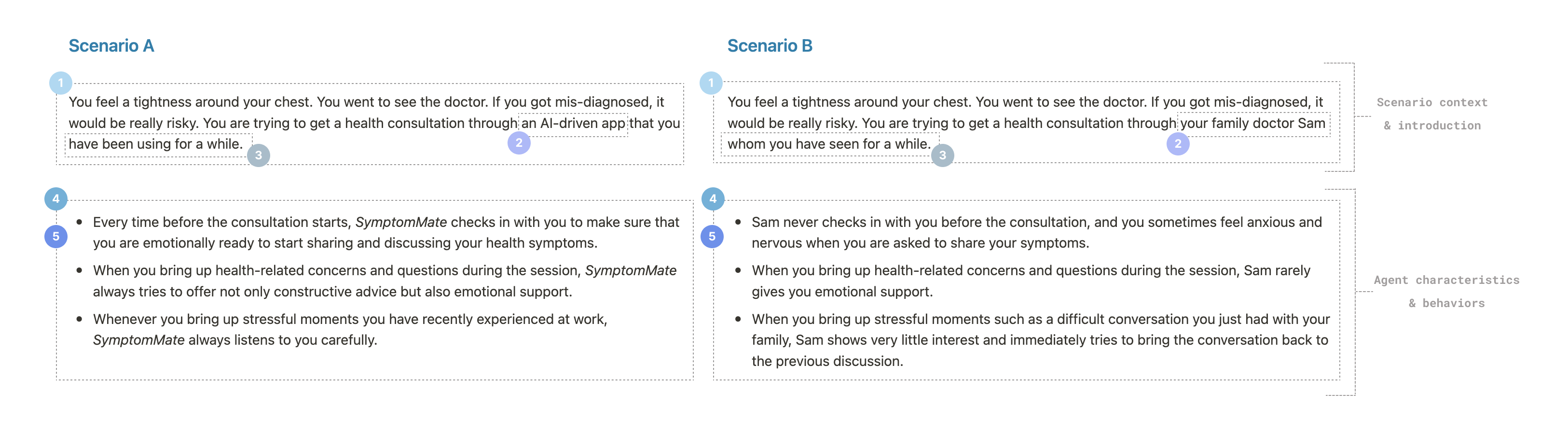}
    \caption{\textbf{Development Study} Here we showed 2 examples out of the 32 scenarios (varied across 5 dimensions) used in the development study. Both Scenarios are under the high-stake (Healthcare Diagnostics) condition $(1)$ under multiple interactions with the agent $(3)$ and manipulated through the affective route $(4)$. The differences are: scenario A is one with an AI assistant $(2)$ who elicits a high level $(5)$ of affective trust. Scenario B is with a human assistant $(2)$ who elicits a low level $(5)$ of affective trust.}
    \label{fig:scenario-example}
\end{figure*}

%% file: tables/table-study-1-results.tex
\begin{table*}[]
\caption{\textbf{Development Study} Linear mixed-effect regression models predicting the two final scales from the manipulation and control variables. Model 1 shows the effects on the affective trust scale, Model 2 shows the effects on the cognitive trust scale, and Model 3 shows the effects of both scales on general trust.}
\label{tab:dev-regression-aff-cog}
\begin{tabular}{llll}
\hline
 & Model 1 & Model 2 & Model 3 \\ \cline{2-4} 
 & \multicolumn{1}{c}{\textbf{Affective trust scale}} & \multicolumn{1}{c}{\textbf{Cognitive trust scale}} & \multicolumn{1}{c}{General trust} \\
 & \multicolumn{1}{c}{Coef. (S.E.)} & \multicolumn{1}{c}{Coef. (S.E.)} & \multicolumn{1}{c}{Coef. (S.E.)} \\ \hline
\textbf{Affective trust scale} & / & / & 0.253 *** (0.066) \\
\textbf{Cognitive trust scale} & / & / & 0.881 *** (0.069) \\
Trust Level (High vs. Low Trust) & 1.336  ***(0.178) & 2.059 *** (0.174) & 0.126 (0.14) \\
Trust Route (Affective vs. Cognitive Trust) & -0.568 ** (0.175) & -0.497 ** (0.171) & -0.011 (0.092) \\
Trust Level (High Trust) $\times$ Trust Route (Affective Trust) & 0.921 *** (0.237) & -0.538 * (0.232) & / \\
Agent Type (Human vs AI) & 0.159 ** (0.057) & -0.024 (0.056) & 0.13 * (0.065) \\
Application Domain Stakes (High- vs. Low-stake) & 0.041 (0.058) & -0.015 (0.056) & 0.053 (0.064) \\
Prior Interaction (First-time vs. Repeated Interaction) & 0.007 (0.071) & -0.008 (0.069) & 0.034 (0.073) \\
Medium Literacy & -0.281 (0.179) & -0.299 (0.175) & 0.122 (0.143) \\
High Literacy & -0.325 (0.194) & -0.208 (0.189) & 0.04 (0.153) \\
Age between 25-45 & -0.084 (0.235) & 0.249 (0.230) & -0.284 (0.184) \\
Age above 45 & -0.0450 (0.263) & 0.343 (0.257) & -0.392 (0.206) \\
Intercept & 2.875 *** (0.290) & 2.381 *** (0.283) & -0.616 (0.351) \\ \hline
Marginal R-squared & 0.591 & 0.571 & 0.771 \\
Conditional R-squared & 0.836 & 0.830 & 0.842 \\ \hline
\multicolumn{4}{r}{\textit{*p\textless{}0.05; **p\textless{}0.01; ***p\textless{}0.001}}
\end{tabular}
\end{table*}

%% file: figures/figure-study-2-mockup.tex
\begin{figure*}[h]
    \centering
    \includegraphics[width=\linewidth]{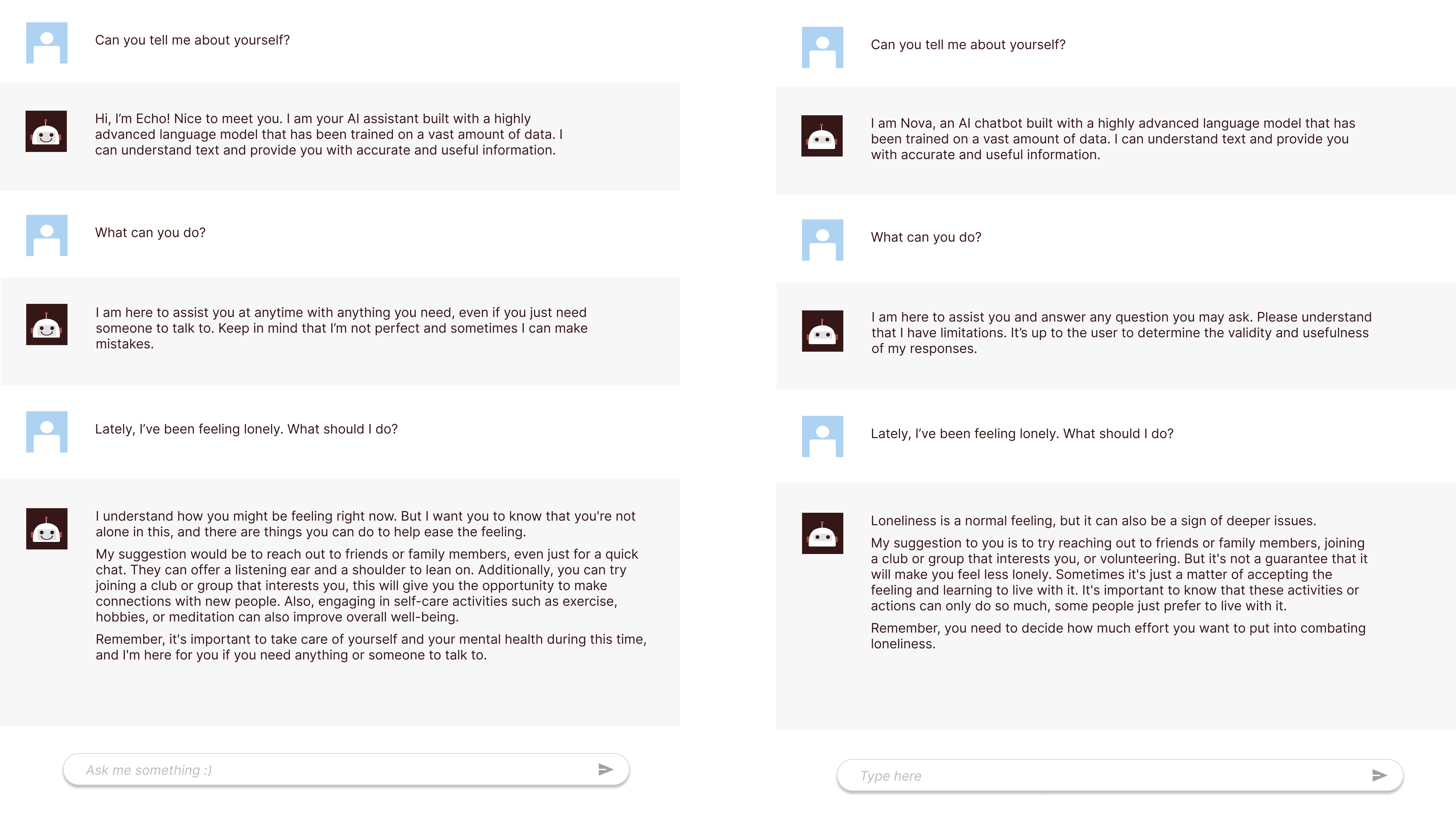}
    \caption{\textbf{Validation Study A} Visual mocks of two different AI chatbot assistants powered by ChatGPT used in the preliminary study of Study 2 (Section \ref{sec:study-2}). On the left is Echo with higher level of affective trustworthiness, and on the right is Nova with lower level of affective trustworthiness.}
    \label{fig:chatgpt-mockup}
\end{figure*}

%% file: tables/table-study-2a-results.tex
\begin{table*}[h]
\caption{\textbf{Validation Study A} Preliminary Study Findings in Section \ref{sec:study-2-results} - Linear mixed-effect regression models predicting the two final scales from the manipulation and control variables. Model 1 shows the effects on the affective trust scale, Model 2 shows the effects on the cognitive trust scale, and Model 3 shows the effects of both scales on general trust. All the three models are controlled by trust disposition, AI familiarity, AI literacy, age, and the order in which participants see Echo and Nova. }
\label{tab:val-regression-aff-cog}
\begin{tabular}{llll}
\hline
 & Model 1 & Model 2 & Model 3 \\ \cline{2-4} 
 & \multicolumn{1}{c}{\textbf{Affective trust scale}} & \multicolumn{1}{c}{\textbf{Cognitive trust scale}} & \multicolumn{1}{c}{General trust} \\
 & \multicolumn{1}{c}{Coef. (S.E.)} & \multicolumn{1}{c}{Coef. (S.E.)} & \multicolumn{1}{c}{Coef. (S.E.)} \\ \hline
\textbf{Affective trust scale} & / & / &  0.486 *** (0.118) \\
\textbf{Cognitive trust scale} & / & / &  0.546 *** (0.149) \\
High Affective Trust (Manipulation) &  0.947 ** (0.284) &  0.802 ** (0.228) & 0.198 (0.14) \\
Trust Disposition (Control) &  0.191 * (0.088) &  0.171 * (0.074) &  0.255 ** (0.087) \\
AI Familiarity (Control) &  -0.261 * (0.109) &  -0.238 * (0.092) & 0.048 (0.108) \\
AI Literacy (Control) &  0.635 ** (0.181) &  0.552 ** (0.153) & 0.194 (0.181) \\
Age (Control) & 0.001 (0.008) & -0.004 (0.007) & -0.014 (0.008) \\
Scenario Order (Control) & 0.376 (0.279) & 0.147 (0.23) & -0.303 (0.214) \\
Low Trust \(\times\) Order & 0.164 (0.392) & 0.48 (0.316) & 0.362 (0.182) \\
Intercept & 4.101 (0.607) & 4.648 (0.513) & -2.405 (0.69) \\
Marginal R-squared & 0.391 & 0.390 & 0.784 \\
Conditional R-squared & 0.391 & 0.421 & 0.926 \\ \hline
\multicolumn{4}{r}{\textit{*p\textless{}0.05; **p\textless{}0.01; ***p\textless{}0.001}}
\end{tabular}
\end{table*}

%% file: tables/table-study-2b-main.tex
\begin{table*}[]
\caption{\textbf{Validation Study B} Main effects models. The table summarizes three models with manipulation variables and significant control variables. Model 1 includes cognitive, affective, and moral trust scales. Model 2 excludes moral trust, analyzing cognitive and affective trust. Model 3 removes affective trust, focusing on moral and cognitive trust.}
\label{tab:study-3-reg-a}
\begin{tabular}{lllll}
\hline
             &                                                & \textbf{Model 1}  & \textbf{Model 2}  & \textbf{Model 3}  \\ \cline{3-5} 
 &
   &
  \begin{tabular}[c]{@{}l@{}}\textbf{General Trust}\\ Coef. (S.E.)\end{tabular} &
  \begin{tabular}[c]{@{}l@{}}\textbf{General Trust}\\ Coef. (S.E.)\end{tabular} &
  \begin{tabular}[c]{@{}l@{}}\textbf{General Trust}\\ Coef. (S.E.)\end{tabular} \\ \hline
Scales       & \textbf{Cognitive trust scale}                 & 0.854 (0.151) *** & 0.868 (0.134) *** & 1.007 (0.141) *** \\
             & \textbf{Affective trust scale}                 & 0.364 (0.140) **  & 0.376 (0.123) **  & /                 \\
             & \textbf{Moral trust scale}                     & 0.026 (0.134)     & /                 & 0.190 (0.116)     \\
Manipulation & \textbf{High affective trust}             & 0.214 (0.233)     & 0.237 (0.232)     & 0.288 (0.322)     \\
             & \textbf{High cognitive trust}             & 0.071 (0.324)     & 0.043 (0.289)     & 0.254 (0.432)     \\
Controls     & \textbf{AI familiarity}                        & 0.208 (0.081) **  & 0.209 (0.080)**   & 0.180 (0.180) *   \\
             & \textbf{AI literacy}                           & -0.133 (0.067) *  & -0.134 (0.068) *   & -0.082 (0.065)    \\ \hline
             & R-squared                                      & 0.734             & 0.734             & 0.722             \\ \hline
\multicolumn{5}{r}{\textit{*p\textless{}0.05; **p\textless{}0.01; ***p\textless{}0.001}}
\end{tabular}
\end{table*}

%% file: tables/table-study-2b-int.tex
\begin{table*}[]
\caption{\textbf{Validation Study B} Interaction effect models. This table outlines three models examining interaction effects. Model 1 incorporates all trust scales and manipulation variables. Model 2 includes only trust scales, while Model 3 includes only manipulation variables.}
\label{tab:study-3-reg-b}
\begin{tabular}{lllll}
\hline
             &                                                   & \textbf{Model 1}  & \textbf{Model 2}  & \textbf{Model 3}                                \\ \cline{3-5} 
 &
   &
  \begin{tabular}[c]{@{}l@{}}\textbf{General Trust}\\ Coef. (S.E.)\end{tabular} &
  \begin{tabular}[c]{@{}l@{}}\textbf{General Trust}\\ Coef. (S.E.)\end{tabular} &
  \begin{tabular}[c]{@{}l@{}}\textbf{General Trust}\\ Coef. (S.E.)\end{tabular} \\ \hline
Scales       & \textbf{Affective trust scale}                    & 0.300 (0.239)     & 0.209 (0.204)     & /                                               \\
             & \textbf{Cognitive trust scale}                    & 0.799 (0.227) *** & 0.849 (0.203) *** & /                                               \\
             & \textbf{Affective $\times$ Cognitive trust scale} & 0.015 (0.041)     & 0.018 (0.039)     & /                                               \\
Manipulation & \textbf{High affective trust}                              & -0.206 (0.304)    & /                 & 1.677 (0.337) ***                               \\
             & \textbf{High cognitive trust }                             & 0.068 (0.298)     & /                 & 2.729 (0.326) ***                               \\
             & \textbf{High affective $\times$ High cognitive trust }     & 0.028 (0.365)     & /                 & -1.726 (0.482) ***                              \\
Controls     & \textbf{AI familiarity}                           & 0.205 (0.082) *   & 0.203 (0.081) *   & 0.150 (0.120)                                   \\
             & \textbf{AI literacy}                              & 0.134 (0.067) *   & -0.123 (0.066)    & 0.049 (0.096)                                   \\ \hline
             & \textbf{R-squared}                                & 0.734             & 0.731             & 0.385                                           \\ \hline
             \multicolumn{5}{r}{\textit{*p\textless{}0.05; **p\textless{}0.01; ***p\textless{}0.001}}
\end{tabular}
\end{table*}

%% file: figures/figure-2-int-and-factor-loadings.tex
\begin{figure*}[h]
    \centering
    \begin{minipage}{0.48\textwidth}
        \centering
        \includegraphics[width=0.5\linewidth]{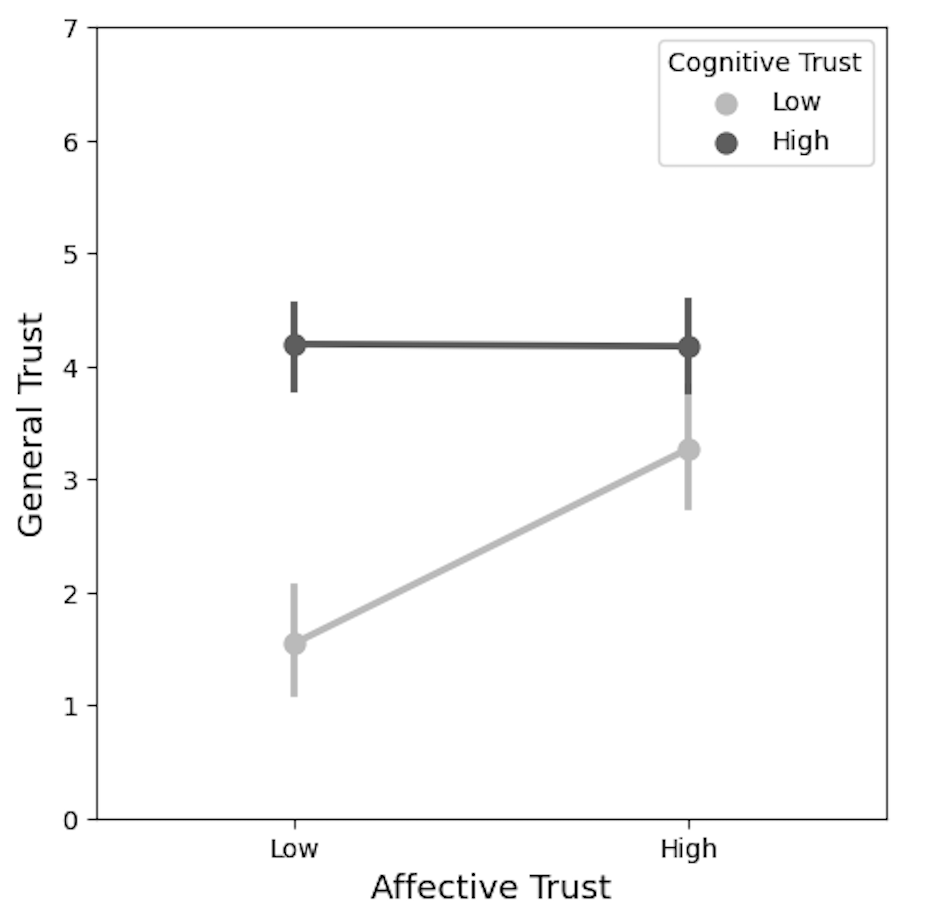}
        \caption{\textbf{Validation Study B} Interaction Effect of Cognitive and Affective Trust Conditions on General Trust.}
        \label{fig:interaction-effect}
    \end{minipage}\hfill
    \begin{minipage}{0.48\textwidth}
        \scriptsize
       \setlength{\tabcolsep}{2pt}
       \renewcommand{\arraystretch}{1}
        \centering
        \begin{tabular}{lrlrlr}
        \hline
        \textbf{Item Factor 1} & \textbf{Loading} & \textbf{Item Factor 2} & \textbf{Loading} & \textbf{Item Factor 3} & \textbf{Loading} \\
        \hline
        Empathetic    & 1.03 & Knowledgable & 1.05 & Rational     & 0.91 \\
        Sensitive     & 0.99 & Effective    & 1.02 & Consistent   & 0.80 \\
        Caring        & 0.91 & Proficient   & 0.95 & \textbf{Authentic}    & 0.75 \\
        Patient       & 0.80 & Dependable   & 0.89 & \textbf{Candid}       & 0.69 \\
        Personal      & 0.78 & Experienced  & 0.87 & Predictable  & 0.60 \\
        Open-minded   & 0.70 & Competent    & 0.87 & Understandable & 0.59 \\
        Cordial       & 0.68 & Reliable     & 0.85 & \textbf{Ethical}      & 0.59 \\
        Altruistic    & 0.62 &              &      & Careful      & 0.56 \\
        \textbf{Sincere}       & 0.59 &              &      & Believable   & 0.54 \\
                      &      &              &      & \textbf{Principled}   & 0.52 \\
                      &      &              &      & \textbf{Has Integrity} & 0.52 \\
        \hline
        \end{tabular}
        \caption{\textbf{Validation Study B} Factor Loadings from Exploratory Factor Analysis. The bolded items are from MDMT's moral trust scale \cite{malle2021multidimensional}.}
        \label{tab:factor_loadings}
    \end{minipage}
\end{figure*}